\documentclass[pdflatex, sn-mathphys, Numbered]{sn-jnl}

\usepackage{graphicx}
\usepackage{multirow}
\usepackage{amsmath,amssymb,amsfonts}
\usepackage{amsthm}
\usepackage{mathrsfs}
\usepackage[title]{appendix}
\usepackage{xcolor}
\usepackage{textcomp}
\usepackage{manyfoot}
\usepackage{booktabs}
\usepackage{algorithm}
\usepackage{algorithmicx}
\usepackage{algpseudocode}
\usepackage{listings}

\usepackage{siunitx}
\DeclareSIUnit\clight{\text{\ensuremath{c}}}

\usepackage{natbib}

\unnumbered

\begin{document}

\title[MMC Array to Study X-ray Transitions in Muonic Atoms]{MMC Array to Study X-ray Transitions in Muonic Atoms}

\author*[1]{\fnm{Daniel} \sur{Unger}}\email{daniel.unger@kip.uni-heidelberg.de}
\author[1]{\fnm{Andreas} \sur{Abeln}}
\author[2]{\fnm{Thomas Elias} \sur{Cocolios}}
\author[3]{\fnm{Ofir} \sur{Eizenberg}}
\author[1]{\fnm{Christian} \sur{Enss}}
\author[1]{\fnm{Andreas} \sur{Fleischmann}}
\author[1]{\fnm{Loredana} \sur{Gastaldo}}
\author[4,5]{\fnm{C\'esar} \sur{Godinho}}
\author[2]{\fnm{Michael} \sur{Heines}}
\author[1]{\fnm{Daniel} \sur{Hengstler}}
\author[5]{\fnm{Paul} \sur{Indelicato}}
\author[1]{\fnm{Ashish} \sur{Jadhav}}
\author[1]{\fnm{Daniel} \sur{Kreuzberger}}
\author[6,7]{\fnm{Klaus} \sur{Kirch}}
\author[7]{\fnm{Andreas} \sur{Knecht}}
\author[4]{\fnm{Jorge} \sur{Machado}}
\author[3]{\fnm{Ben} \sur{Ohayon}}
\author[5]{\fnm{Nancy} \sur{Paul}}
\author[8,10]{\fnm{Randolf} \sur{Pohl}}
\author[6,7]{\fnm{Katharina} \sur{von~Schoeler}}
\author[7]{\fnm{Stergiani Marina} \sur{Vogiatzi}}
\author[9,10]{\fnm{Frederik} \sur{Wauters}}

\affil*[1]{\orgdiv{Kirchhoff‐Institut für Physik}, \orgname{Universität Heidelberg}, \orgaddress{\street{Im Neuenheimer Feld 227}, \postcode{69120} \city{Heidelberg}, \country{Germany}}}
\affil[2]{\orgdiv{Instituut voor Kern- en Stralingsfysica}, \orgname{KU Leuven}, \orgaddress{\street{Celestijnenlaan 200d}, \postcode{3001} \city{Leuven}, \country{Belgium}}}
\affil[3]{\orgdiv{Physics Department}, \orgname{Technion -- Israel Institute of Technology}, \orgaddress{\postcode{32000} \city{Haifa}, \country{Israel}}}
\affil[4]{\orgdiv{Laboratory of Instrumentation, Biomedical Engineering and Radiation Physics, Department of Physics, NOVA School of Science and Technology}, \orgname{NOVA University Lisbon}, \orgaddress{\postcode{2829-516} \city{Caparica}, \country{Portugal}}}
\affil[5]{\orgdiv{Laboratoire Kastler Brossel}, \orgname{Sorbonne Université, CNRS, ENS-Université PSL, Collège de France, Campus Pierre et Marie Curie}, \orgaddress{\street{Case 74, 4, Place Jussieu}, \postcode{75005} \city{Paris}, \country{France}}}
\affil[6]{\orgdiv{Institute for Particle Physics and Astrophysics}, \orgname{ETH Zürich}, \orgaddress{\street{Otto-Stern-Weg 5}, \postcode{8093} \city{Zürich}, \country{Switzerland}}}
\affil[7]{\orgname{Paul Scherrer Institut}, \orgaddress{\street{Forschungsstrasse 111}, \postcode{5232}~\city{Villigen}, \country{Switzerland}}}
\affil[8]{\orgdiv{Institut für Physik, QUANTUM}, \orgname{Johannes Gutenberg-Universität Mainz}, \orgaddress{\postcode{55128} \city{Mainz}, \country{Germany}}}
\affil[9]{\orgdiv{Institut für Kernphysik}, \orgname{Johannes Gutenberg-Universität Mainz}, \orgaddress{\postcode{55128} \city{Mainz}, \country{Germany}}}
\affil[10]{\orgdiv{PRISMA+ Cluster of Excellence}, \orgname{Johannes Gutenberg-Universität Mainz}, \orgaddress{\postcode{55128} \city{Mainz}, \country{Germany}}}

\abstract{The QUARTET collaboration aims to significantly improve the precision of the absolute nuclear charge radii of light nuclei from Li to Ne by using an array of metallic magnetic calorimeters to perform high-precision X-ray spectroscopy of low-lying states in muonic atoms. A proof-of-principle measurement with lithium, beryllium and boron is planned for fall 2023 at the Paul Scherrer Institute. We discuss the performance achieved with the maXs-30 detector module to be used. To place the detector close to the target chamber where the muon beam will impact the material under study, we have developed a new dilution refrigerator sidearm. We further discuss the expected efficiency given the transparency of the X-ray windows and the quantum efficiency of the detector. The expected muonic X-ray rate combined with the high resolving power and detection efficiency of the detector suggest that QUARTET will be able to study the de-excitation of light muonic atoms at an unprecedented level, increasing the relative energy resolution by up to a factor of \num{20} compared to conventional detector techniques.}

\keywords{Metallic Magnetic Calorimeter, X-ray Spectroscopy, Muonic Atoms, Exotic Atoms, Absolute Nuclear Charge Radii, QUARTET}

\maketitle

\section{Introduction}\label{sec1}

In a muonic atom, at least one of the electrons is replaced by a muon. The energy levels of the system are strongly influenced by nuclear finite-size effects due to the strong overlap between the wave functions of the muon and the nucleus. Spectroscopy of muonic atom transitions can therefore provide information about the nuclear structure. The QUARTET (QUAntum inteRacTions with Exotic aToms) collaboration aims for high-resolution spectroscopy of light muonic atoms from lithium to neon. We intend to improve the precision of the absolute nuclear charge radii for these elements by factors of \numrange[range-phrase = --]{3}{20}. For detailed information on the physics cases and theoretical considerations, we refer the reader to an accompanying publication~\cite{Ohayon_2023}.

At the Paul Scherrer Institute (PSI) in Switzerland, beam time has been granted for a proof-of-principle measurement in the fall of 2023, where as a first step we will attempt to measure the transition energies in muonic lithium, beryllium, and boron. In addition, we want to observe the isotope shift in the muonic transitions between $^6$Li and $^7$Li as well as between $^{10}$B and $^{11}$B. A muon beam with a rate of \SI{10}{\kilo\hertz} and a momentum of \SI{28}{\mega\electronvolt\per\clight} directed at an exchangeable target inside a vacuum chamber is used to create muonic atoms. The muon is captured at a high principal quantum number $n$ and cascades toward the ground state of the atom by emitting X-rays before decaying or being captured by the nucleus~\cite{Measday_2001}. The most prominent muonic X-ray lines are in the energy range between \SI{10}{\kilo\electronvolt} and \SI{60}{\kilo\electronvolt}.

We plan to use a low-temperature X-ray detector based on metallic magnetic calorimeter (MMC)~\cite{Hengstler_2015, Unger_2021} to measure the X-rays emitted by the muonic atoms. These highly linear detectors feature high resolving power over a wide energy range~\cite{Fleischmann_2005}, providing a suitable technology for high-precision X-ray spectroscopy of light muonic atoms. For efficient placement of the relatively small detector close to the muon target, the detector will be integrated into a new refrigerator sidearm. In addition, existing muon, electron, and photon detectors from the muX experiment will be used for monitoring, validation and background suppression~\cite{Wauters_2021}. In the following we will discuss the MMC-based detector, the new refrigerator sidearm, and the expected performance in more detail.

\section{Detector Module for QUARTET}\label{sec2}

An MMC is a low-temperature micro-calorimeter based on a paramagnetic temperature sensor~\cite{Fleischmann_2005}, typically operated in a dilution refrigerator at \SI{20}{\milli\kelvin} and read-out using a two-stage SQUID (superconducting quantum interference device) circuit~\cite{Kempf_2015, Mantegazzini_2021}. MMCs are used in a wide variety of experiments~\cite{Gastaldo_2017, Unger_2021, Herdrich_2023}: They are characterized by high resolving power up to \num{6000} over a wide energy range~\cite{Kempf_2018, Sikorsky_2020} while being highly linear with a well-understood energy calibration~\cite{Hengstler_2015} and having a low energy threshold below \SI{100}{\electronvolt}~\cite{Ranitzsch_2017, Unger_2021}. They can be designed to have a high quantum efficiency in the energy range of interest up to nearly \SI{100}{\percent}~\cite{Fleischmann_2009}, and can have a fast intrinsic response time up to \SI{100}{\nano\second}~\cite{Pies_2012}. This makes MMCs very attractive for high-resolution X-ray spectroscopy.

An MMC has a particle absorber designed to stop incoming particles of interest and a paramagnetic temperature sensor in thermal contact with the absorber. After a particle interacts with the absorber, the temperature of the detector increases proportionally to the absorbed energy. The sensor, located in a weak static magnetic field, has a temperature-dependent magnetization. The change of magnetization is measured as a change of magnetic flux, which is coupled to the input coil of a dc-SQUID with a superconducting coil. Thus, the screening current generated in the superconducting coil induces a flux change in the SQUID, resulting in a voltage change proportional to the energy of the incident particle.

\begin{figure}[htb]
\centering
\includegraphics[width=4in]{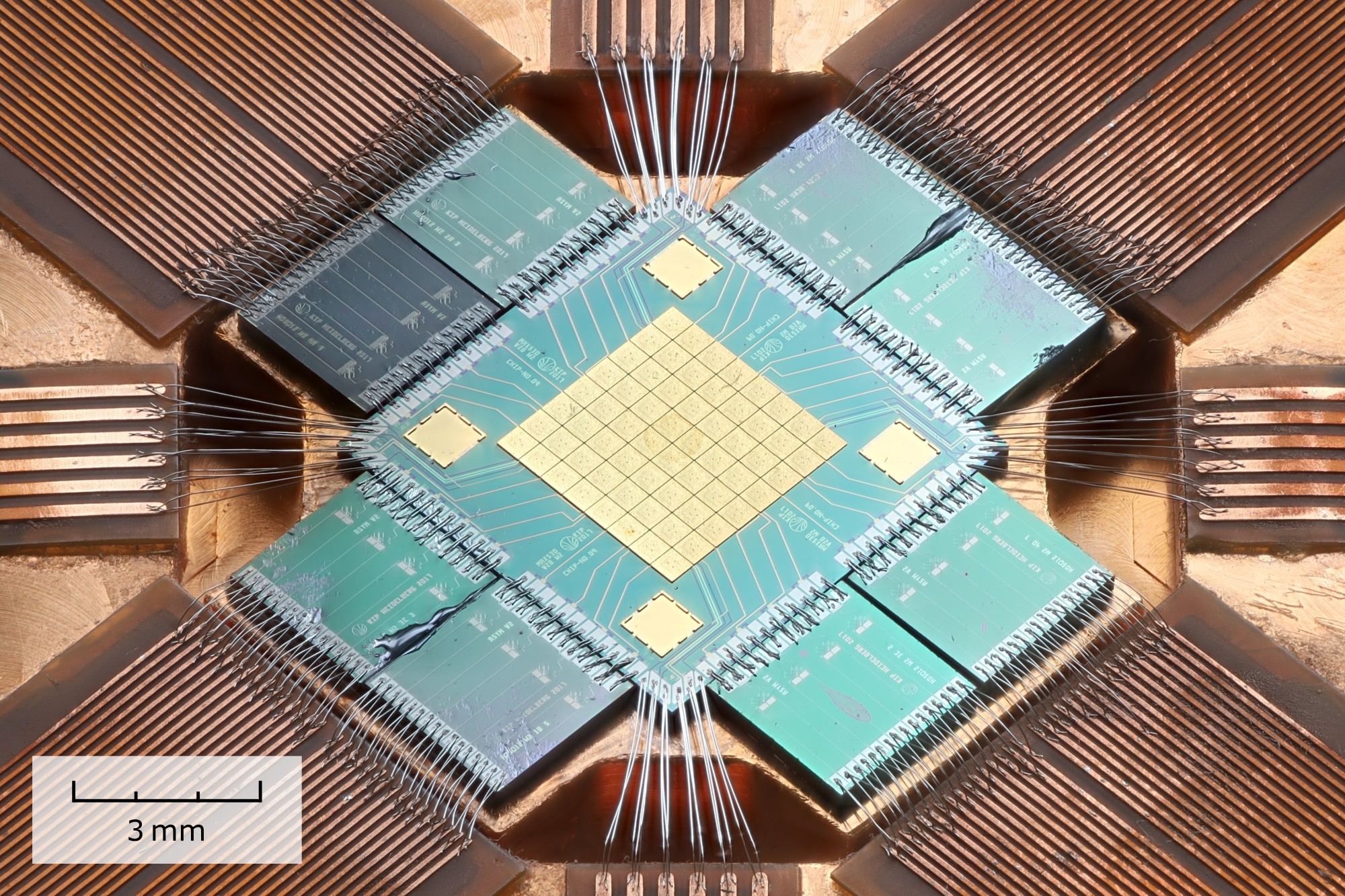}
\caption{The maXs-30 detector and eight first-stage SQUID chips of the detector module to be used for the proof-of-principle measurement at PSI~\cite{Unger_2021}.}\label{fig1}
\end{figure}

The X-ray detector which we plan to use for the first QUARTET measurement is a maXs-30 (micro-calorimeter array for X-ray spectroscopy) originally designed for experiments at the heavy ion storage ring ESR at the GSI accelerator~\cite{Hengstler_2015}. The 64-pixel two-dimensional detector array has a total active area of \SI{4}{\milli\meter}~$\times$~\SI{4}{\milli\meter} and is optimized for X-ray spectroscopy up to \SI{30}{\kilo\electronvolt} with its \SI{20}{\micro\meter} thick gold absorbers. The detector is read out by 32 individual two-stage SQUID channels, where a single channel reads out two pixels of gradiometric design with signals of opposite polarity~\cite{Fleischmann_2009}. The simulated FWHM baseline energy resolution given the expected noise and signal height is $\SI{6}{\electronvolt}$ at a temperature of \SI{20}{\milli\kelvin}. The maXs-30 detector module to be used was developed and characterized as part of the detector development for IAXO \cite{Unger_2021}. The module, shown in figure~\ref{fig1}, has 62 out of 64 operational pixels, shows a very linear response of \SI{0.1}{\percent} at \SI{6}{\kilo\electronvolt} and achieves a mean energy resolution of \SI{6}{\electronvolt} FWHM at a detector temperature of \SI{17}{\milli\kelvin}~\cite{Unger_2021}. With the most prominent X-ray lines from the muonic atoms of interest in the energy range between \SI{10}{\kilo\electronvolt} and \SI{60}{\kilo\electronvolt}, the module is well suited for the proof-of-principle measurement.

\section{Dilution Refrigerator Sidearm}\label{sec3}

\begin{figure}[htb]
\centering
\includegraphics[width=4in]{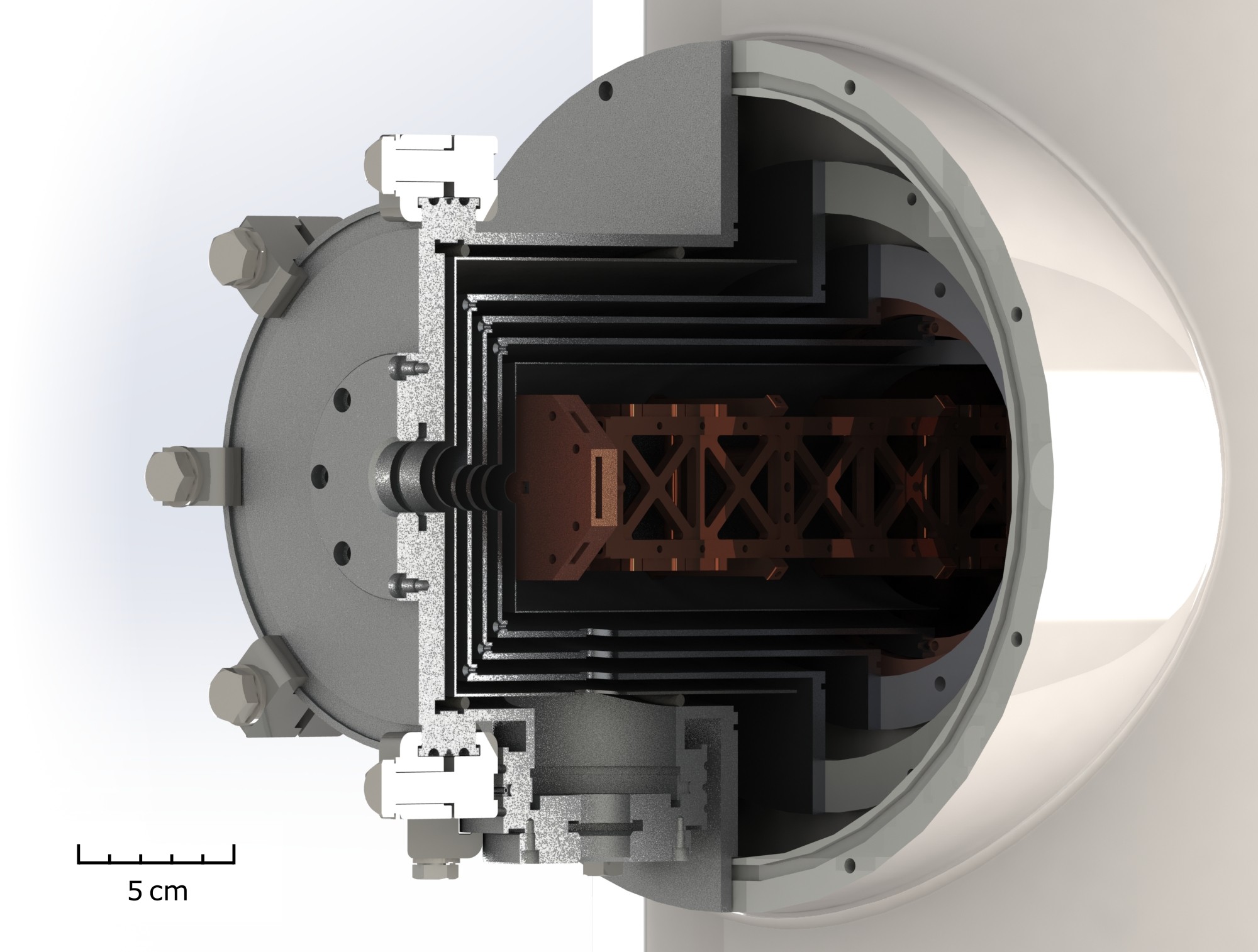}
\caption{Design of the new sidearm for the dilution refrigerator. Multiple shields protect the detector from thermal radiation and magnetic field changes. The sidearm can be rotated and has two standard vacuum flanges. X-ray windows allow the measurement of external X-rays.}\label{fig2}
\end{figure}

For efficient integration of the detector module to the muon target chamber, we have designed a new sidearm for a commercially available dilution refrigerator from Bluefors\footnote{https://bluefors.com/}, shown in figure~\ref{fig2}. The sidearm consists of an aluminum vacuum shield with an attached Mu-metal shield, three aluminum thermal shields (connected to the \SI{50}{\kelvin}, \SI{4}{\kelvin}, and still temperature stages in the refrigerator), and an extension arm made out of a copper structure covered by a cylindrical niobium shield. The copper structure is attached to the bottom of the mixing chamber plate of the dilution refrigerator and can hold a detector module up to a cylindrical shaped volume of \SI{8}{\centi\meter} in diameter and height. The mounting on the mixing chamber allows for a fine alignment in all three axes of at least \SI{3}{\milli\meter} to align with respect to the outer shields. The different expansion coefficients of the materials have been taken into account so that the shields align when the detector is cold. The sidearm can be rotated to allow for different detector orientations. Two standard blind flanges on the vacuum shield can be replaced or removed for connection to other vacuum systems.

The vacuum shield and thermal shields are made of pure aluminum to minimize the amount of magnetic material near the detector. The Mu-metal, still, and niobium shields protect the detector from external magnetic fields and magnetic field changes. Once the side arm is installed and aligned, the entire inner copper frame with the prepared detector module can be removed as one piece and quickly reinstalled later. The thermal shields are fabricated to fit the refrigerator and require no further alignment. Quick assembly of the sidearm and its shields is critical when preparing for a measurement with very limited beam time.

X-ray windows are used to allow external X-rays to enter and reach the detector. We plan to use an AP5 window from Moxtek with a diameter of \SI{14}{\milli\meter} on the vacuum shield, while a total of three \SI{6}{\micro\meter} thick Mylar foils coated on both sides with \SI{40}{\nano\meter} aluminum are planned for the three inner thermal shields. The detector mounted on the extension arm surpasses the mixing chamber plate by roughly \SI{20}{\centi\meter}. The solid angle aperture of the detector is about \num{6e-2}, limited by the outermost X-ray window on the vacuum shield, which is about \SI{5}{\centi\meter} further away. The vacuum shield of the sidearm extends beyond the refrigerator frame, making it the outermost part and allowing it to be positioned very close to the target chamber while leaving the necessary space to integrate germanium detectors, silicon detectors, and scintillators for monitoring, validation, and background suppression.

\section{Expected Performance}\label{sec4}

\begin{figure}[htb]
\centering
\includegraphics[width=5in]{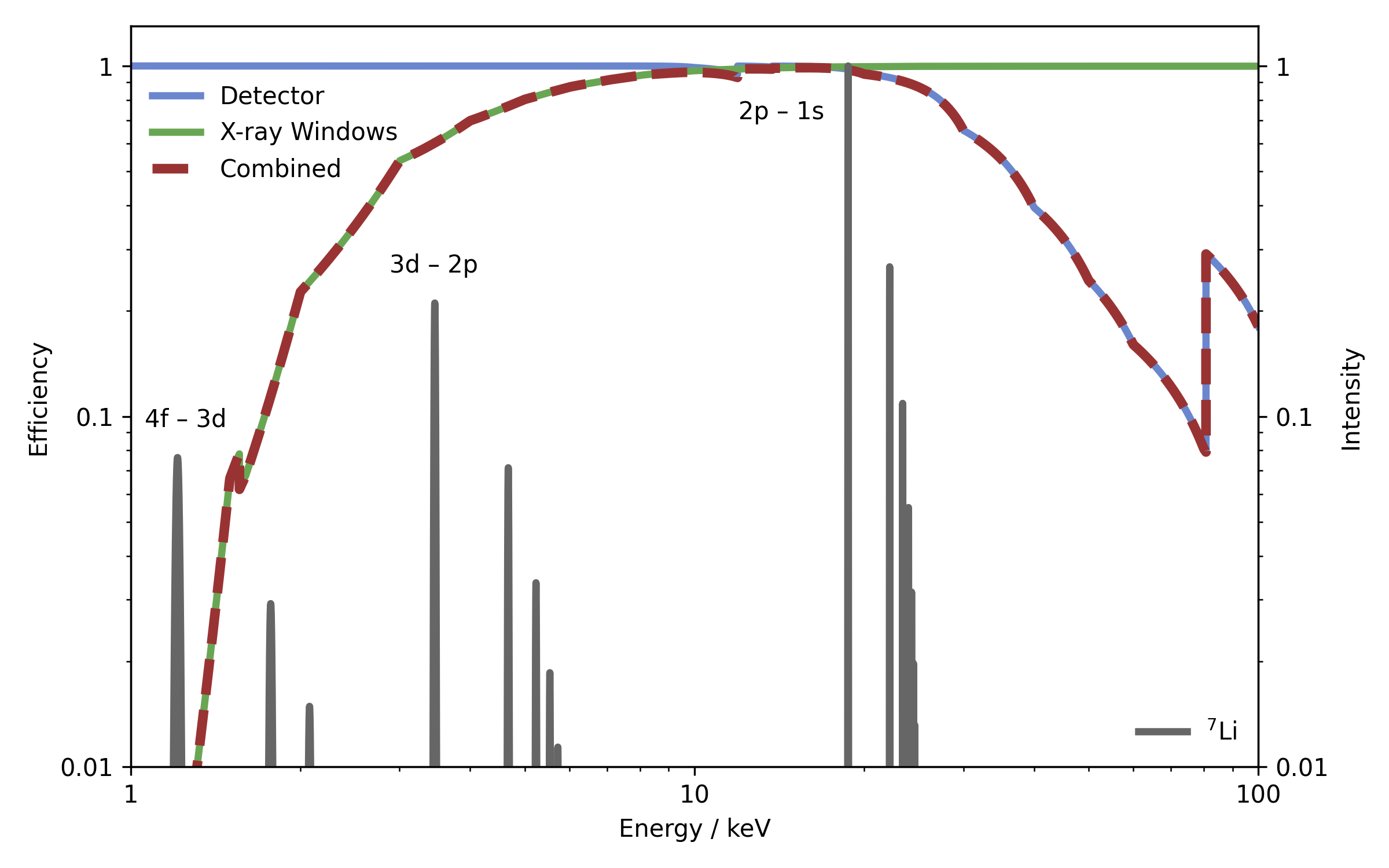}
\caption{Expected efficiency considering the quantum efficiency of the detector (blue) and the transmission of the X-ray windows (green). The combined efficiency (red-brown, dashed) is above \SI{20}{\percent} between \num{2} and \SI{50}{\kilo\electronvolt}. As an example, a calculated spectrum with the roughly estimated energies and intensities of the most prominent muonic $^7$Li lines is also shown, considering a finite detector response of \SI{10}{\electronvolt} FWHM (gray). For the estimation, attenuation coefficients were taken from NIST~\cite{NIST_2004}, and the efficiency of the AP5 X-ray window was provided by Moxtek.}\label{fig3}
\end{figure}

The expected efficiency of the setup considering the aforementioned X-ray windows and the absorber thickness of the MMCs is shown in figure~\ref{fig3}. The combined efficiency is reduced by the X-ray windows absorption below \SI{10}{\kilo\electronvolt} and by the absorber thickness of the MMCs above \SI{20}{\kilo\electronvolt}. The quantum efficiency of the absorber is above \SI{50}{\percent} up to \SI{40}{\kilo\electronvolt}. The efficiency of the available X-ray window configuration is strongly dominated by the relatively thick Mylar foils. In the future, the thickness could be reduced to about \SI{1}{\micro\meter}, increasing the efficiency at low energies by more than an order of magnitude to \SI{20}{\percent} at \SI{1}{\kilo\electronvolt}.

As an example, the expected muonic $^7$Li spectrum is shown in figure~\ref{fig3}. The most prominent line from the 2p--1s transition is in the very high-efficiency region. At \SI{19}{\kilo\electronvolt} we anticipate an efficiency of \SI{97}{\percent}. We also estimate reasonable efficiencies for the 2p--1s line of a beryllium or boron target (\SI{59}{\percent} at \SI{33}{\kilo\electronvolt} for beryllium and \SI{22}{\percent} at \SI{53}{\kilo\electronvolt} for boron). The muon target will be about \SI{15}{\centi\meter} away from the MMC detector array. Given the solid angle covered and the expected muon rate (\SI{1e4}{\per\second} triggered events), a rough estimate gives an expected rate on the order of \SI{0.3}{\per\second} muon-induced X-ray events, with roughly \SI{0.15}{\per\second} events at the MMC array from the 2p--1s transition, which is well below the maximum rate that the detector can reliably handle. Improving the radii by an order of magnitude requires a ppm precision on the line position. For the muonic 2p--1s transition of $^7$Li at \SI{19}{\kilo\electronvolt}, this corresponds to a precision of \SI{0.1}{\electronvolt}. For a detector with an FWHM energy resolution of \SI{10}{\electronvolt}, a few thousand events are sufficient to achieve the required statistical precision. Given the expected rate of \SI{0.15}{\per\second}, this could be achieved with a few hours of measurement.

\section{Conclusion}\label{sec5}

The QUARTET collaboration aims at performing high-precision X-ray spectroscopy of light muonic atoms from lithium to neon using a two-dimensional MMC array. For a proof-of-principle measurement with lithium, beryllium, and boron targets, we have prepared a new dedicated experimental setup with an already well characterized detector module. The main feature of this setup is a new refrigerator sidearm that allows for more efficient detector placement close to the muon target. We expect a suitable rate for de-excitation spectra for light muonic atoms. Combined with the excellent performance of the MMC array, we could exceed the current accuracy on the position of muonic X-ray lines already with the scheduled proof-of-principle measurement at PSI in October 2023. This will be the first use of MMCs to study exotic atoms.

\backmatter

\section*{Acknowledgments}

We are grateful to the Max Planck Institute for Nuclear Physics in Heidelberg, especially the precision mechanics team, for their support in the fabrication of the refrigerator sidearm in cooperation with the precision mechanics team at the Kirchhoff Institute for Physics in Heidelberg. We thank the cleanroom team at the Kirchhoff Institute for Physics in Heidelberg for their contribution to the fabrication of the used detector and SQUIDs.

\section*{Funding}

The group at the Kirchhoff Institute for Physics is supported by the Field of Focus II Initiative at Heidelberg University funded by the "Bundesministerium für Bildung und Forschung" (BMBF) and by the "Wissenschaftsministerium Baden-Württemberg" in the framework of the "Exzellenzstrategie von Bund und Ländern". DU acknowledges the support of the Research Training Group HighRR (GRK 2058) funded by the Deutsche Forschungsgemeinschaft (DFG). TEC and MH acknowledge the support of the FWO-Vlaanderen (Belgium) and of KU Leuven (C14/22/104).

\section*{Declarations}

The authors have no conflicts of interest to declare that are relevant to the content of this article.

\bibliography{bibliography}

%% BioMed_Central_Bib_Style_v1.01

\begin{thebibliography}{14}
% BibTex style file: bmc-mathphys.bst (version 2.1), 2014-07-24
\ifx \bisbn   \undefined \def \bisbn  #1{ISBN #1}\fi
\ifx \binits  \undefined \def \binits#1{#1}\fi
\ifx \bauthor  \undefined \def \bauthor#1{#1}\fi
\ifx \batitle  \undefined \def \batitle#1{#1}\fi
\ifx \bjtitle  \undefined \def \bjtitle#1{#1}\fi
\ifx \bvolume  \undefined \def \bvolume#1{\textbf{#1}}\fi
\ifx \byear  \undefined \def \byear#1{#1}\fi
\ifx \bissue  \undefined \def \bissue#1{#1}\fi
\ifx \bfpage  \undefined \def \bfpage#1{#1}\fi
\ifx \blpage  \undefined \def \blpage #1{#1}\fi
\ifx \burl  \undefined \def \burl#1{\textsf{#1}}\fi
\ifx \doiurl  \undefined \def \doiurl#1{\url{https://doi.org/#1}}\fi
\ifx \betal  \undefined \def \betal{\textit{et al.}}\fi
\ifx \binstitute  \undefined \def \binstitute#1{#1}\fi
\ifx \binstitutionaled  \undefined \def \binstitutionaled#1{#1}\fi
\ifx \bctitle  \undefined \def \bctitle#1{#1}\fi
\ifx \beditor  \undefined \def \beditor#1{#1}\fi
\ifx \bpublisher  \undefined \def \bpublisher#1{#1}\fi
\ifx \bbtitle  \undefined \def \bbtitle#1{#1}\fi
\ifx \bedition  \undefined \def \bedition#1{#1}\fi
\ifx \bseriesno  \undefined \def \bseriesno#1{#1}\fi
\ifx \blocation  \undefined \def \blocation#1{#1}\fi
\ifx \bsertitle  \undefined \def \bsertitle#1{#1}\fi
\ifx \bsnm \undefined \def \bsnm#1{#1}\fi
\ifx \bsuffix \undefined \def \bsuffix#1{#1}\fi
\ifx \bparticle \undefined \def \bparticle#1{#1}\fi
\ifx \barticle \undefined \def \barticle#1{#1}\fi
\bibcommenthead
\ifx \bconfdate \undefined \def \bconfdate #1{#1}\fi
\ifx \botherref \undefined \def \botherref #1{#1}\fi
\ifx \url \undefined \def \url#1{\textsf{#1}}\fi
\ifx \bchapter \undefined \def \bchapter#1{#1}\fi
\ifx \bbook \undefined \def \bbook#1{#1}\fi
\ifx \bcomment \undefined \def \bcomment#1{#1}\fi
\ifx \oauthor \undefined \def \oauthor#1{#1}\fi
\ifx \citeauthoryear \undefined \def \citeauthoryear#1{#1}\fi
\ifx \endbibitem  \undefined \def \endbibitem {}\fi
\ifx \bconflocation  \undefined \def \bconflocation#1{#1}\fi
\ifx \arxivurl  \undefined \def \arxivurl#1{\textsf{#1}}\fi
\csname PreBibitemsHook\endcsname

%%% 1
\bibitem[\protect\citeauthoryear{Ohayon and \textit{et~al.}}{2023}]{Ohayon_2023}
\begin{botherref}
\oauthor{\bsnm{Ohayon}, \binits{B.}},
\oauthor{\bsnm{\textit{et~al.}}}:
Towards precision muonic x-ray measurements of charge radii of light nuclei
(2023)
{\href{https://arxiv.org/abs/2310.03846}{{arXiv:2310.03846}}}
{[physics.atom-ph]}
\end{botherref}
\endbibitem

%%% 2
\bibitem[\protect\citeauthoryear{Measday}{2001}]{Measday_2001}
\begin{barticle}
\bauthor{\bsnm{Measday}, \binits{D.F.}}:
\batitle{The nuclear physics of muon capture}.
\bjtitle{Phys. Rep.}
\bvolume{354}(\bissue{4}),
\bfpage{243}--\blpage{409}
(\byear{2001})
\doiurl{10.1016/S0370-1573(01)00012-6}
\end{barticle}
\endbibitem

%%% 3
\bibitem[\protect\citeauthoryear{Hengstler et~al.}{2015}]{Hengstler_2015}
\begin{barticle}
\bauthor{\bsnm{Hengstler}, \binits{D.}}, \betal:
\batitle{Towards {FAIR}: first measurements of metallic magnetic calorimeters for high-resolution {X}-ray spectroscopy at {GSI}}.
\bjtitle{Phys. Scr.}
\bvolume{2015}(\bissue{T166}),
\bfpage{014054}
(\byear{2015})
\doiurl{10.1088/0031-8949/2015/T166/014054}
\end{barticle}
\endbibitem

%%% 4
\bibitem[\protect\citeauthoryear{Unger et~al.}{2021}]{Unger_2021}
\begin{barticle}
\bauthor{\bsnm{Unger}, \binits{D.}}, \betal:
\batitle{High-resolution for {IAXO}: {MMC}-based {X}-ray detectors}.
\bjtitle{JINST}
\bvolume{16}(\bissue{06}),
\bfpage{06006}
(\byear{2021})
\doiurl{10.1088/1748-0221/16/06/P06006}
\end{barticle}
\endbibitem

%%% 5
\bibitem[\protect\citeauthoryear{Fleischmann et~al.}{2005}]{Fleischmann_2005}
\begin{bbook}
\bauthor{\bsnm{Fleischmann}, \binits{A.}}, \betal:
In: \beditor{\bsnm{Enss}, \binits{C.}} (ed.)
\bbtitle{Metallic Magnetic Calorimeters},
pp. \bfpage{151}--\blpage{216}.
\bpublisher{Springer},
\blocation{Berlin, Heidelberg}
(\byear{2005}).
\doiurl{10.1007/10933596_4}
\end{bbook}
\endbibitem

%%% 6
\bibitem[\protect\citeauthoryear{Wauters et~al.}{2021}]{Wauters_2021}
\begin{barticle}
\bauthor{\bsnm{Wauters}, \binits{F.}}, \betal:
\batitle{The {muX} project}.
\bjtitle{SciPost Phys. Proc.}
\bvolume{5},
\bfpage{022}
(\byear{2021})
\doiurl{10.21468/SciPostPhysProc.5.022}
\end{barticle}
\endbibitem

%%% 7
\bibitem[\protect\citeauthoryear{Kempf et~al.}{2015}]{Kempf_2015}
\begin{barticle}
\bauthor{\bsnm{Kempf}, \binits{S.}}, \betal:
\batitle{Direct-current superconducting quantum interference devices for the readout of metallic magnetic calorimeters}.
\bjtitle{Supercond. Sci. Technol.}
\bvolume{28}(\bissue{4}),
\bfpage{045008}
(\byear{2015})
\doiurl{10.1088/0953-2048/28/4/045008}
\end{barticle}
\endbibitem

%%% 8
\bibitem[\protect\citeauthoryear{Mantegazzini et~al.}{2021}]{Mantegazzini_2021}
\begin{barticle}
\bauthor{\bsnm{Mantegazzini}, \binits{F.}}, \betal:
\batitle{Multichannel read-out for arrays of metallic magnetic calorimeters}.
\bjtitle{JINST}
\bvolume{16}(\bissue{08}),
\bfpage{08003}
(\byear{2021})
\doiurl{10.1088/1748-0221/16/08/P08003}
\end{barticle}
\endbibitem

%%% 9
\bibitem[\protect\citeauthoryear{Gastaldo et~al.}{2017}]{Gastaldo_2017}
\begin{barticle}
\bauthor{\bsnm{Gastaldo}, \binits{L.}}, \betal:
\batitle{The electron capture in {$^{163}\mathrm{Ho}$} experiment -- {ECHo}}.
\bjtitle{Eur. Phys. J. Spec. Top.}
\bvolume{226}(\bissue{8}),
\bfpage{1623}--\blpage{1694}
(\byear{2017})
\doiurl{10.1140/epjst/e2017-70071-y}
\end{barticle}
\endbibitem

%%% 10
\bibitem[\protect\citeauthoryear{Sikorsky et~al.}{2020}]{Sikorsky_2020}
\begin{barticle}
\bauthor{\bsnm{Sikorsky}, \binits{T.}}, \betal:
\batitle{Measurement of the {$^{229}\mathrm{Th}$} isomer energy with a magnetic microcalorimeter}.
\bjtitle{Phys. Rev. Lett.}
\bvolume{125},
\bfpage{142503}
(\byear{2020})
\doiurl{10.1103/PhysRevLett.125.142503}
\end{barticle}
\endbibitem

%%% 11
\bibitem[\protect\citeauthoryear{Kempf et~al.}{2018}]{Kempf_2018}
\begin{barticle}
\bauthor{\bsnm{Kempf}, \binits{S.}}, \betal:
\batitle{Physics and applications of metallic magnetic calorimeters}.
\bjtitle{J. Low Temp. Phys.}
\bvolume{193}(\bissue{3}),
\bfpage{365}--\blpage{379}
(\byear{2018})
\doiurl{10.1007/s10909-018-1891-6}
\end{barticle}
\endbibitem

%%% 12
\bibitem[\protect\citeauthoryear{Fleischmann et~al.}{2009}]{Fleischmann_2009}
\begin{barticle}
\bauthor{\bsnm{Fleischmann}, \binits{A.}}, \betal:
\batitle{Metallic magnetic calorimeters}.
\bjtitle{AIP Conf. Proc.}
\bvolume{1185}(\bissue{1}),
\bfpage{571}--\blpage{578}
(\byear{2009})
\doiurl{10.1063/1.3292407}
\end{barticle}
\endbibitem

%%% 13
\bibitem[\protect\citeauthoryear{Pies et~al.}{2012}]{Pies_2012}
\begin{barticle}
\bauthor{\bsnm{Pies}, \binits{C.}}, \betal:
\batitle{{maXs}: Microcalorimeter arrays for high-resolution {X}-ray spectroscopy at {GSI/FAIR}}.
\bjtitle{J. Low Temp. Phys.}
\bvolume{167}(\bissue{3}),
\bfpage{269}--\blpage{279}
(\byear{2012})
\doiurl{10.1007/s10909-012-0557-z}
\end{barticle}
\endbibitem

%%% 14
\bibitem[\protect\citeauthoryear{Hubbell and Seltzer}{2004}]{NIST_2004}
\begin{botherref}
\oauthor{\bsnm{Hubbell}, \binits{J.H.}},
\oauthor{\bsnm{Seltzer}, \binits{S.M.}}:
Tables of {X}-ray mass attenuation coefficients and mass energy-absorption coefficients
(2004)
\doiurl{10.18434/T4D01F}
\end{botherref}
\endbibitem

\end{thebibliography}

\end{document}